\documentclass[12pt]{article}
\usepackage{graphicx}
\usepackage{booktabs}
\usepackage{float}
\usepackage[margin=1in,letterpaper]{geometry}
\usepackage{xcolor}

\begin{document}

\author{ A. Saleh, M. I. Jaghoub\thanks
{E-mail address: mjaghoub@ju.edu.jo}   \\ [1ex]
\small{Physics Department, The University of Jordan, P.C. 11942,	Amman, Jordan}   }
\title{Modeling alpha-nucleus elastic scattering using a velocity-dependent optical model } 
\maketitle

\begin{abstract}

We performed a least-square fit analysis to reproduce the elastic angular distributions for $\alpha$ scattering on various nuclei form $^{12}$C to $^{208}$Pb for incident energies in the range 18 - 70 MeV using a velocity-dependent optical model. The model reproduced the experimental data well including the enhanced angular distributions in the large-angle scattering region, which is commonly known as the anomalous large angle scattering (ALAS). Our best-fit potential parameters are linear functions of incident energy. Although the ALAS effect is not present in the case of $\alpha$ scattering on the intermediate $^{58}$Ni and heavy $^{208}$Pb nuclei, we considered these nuclei to demonstrate the effectiveness of the VDOM in describing the angular distributions for $\alpha$ scattering on various light, intermediate and heavy nuceli. For $\alpha$ scattering on $^{40}$Ca, we compared our results to two previous works that adopted the conventional optical model. One model reproduced the data better at low energies, while the other performed better at high energies. In contrast, the velocity-dependent model of this work described the data across the considered angular range. 
\color{black}
\end{abstract}

\section{Introduction}

The scattering of $\alpha-$particles on light and intermediate nuclei has been widely investigated. In the forward direction, the elastic angular distributions show the expected diffraction pattern.  However, the distributions are enhanced in the back-angle scattering region ($\theta > 90^0$), which is commonly known as anomalous large angle scattering (ALAS) \cite{PhysRevC.7.1368}, \cite{BOBROWSKA1969361}, \cite{EBERHARD1970343}, \cite{PLANETA197997} and \cite{GRUHN1966161} and references there in. Several attempts, based on different explanations, have been made to explain the observed ALAS effect. For example, a detailed optical model analysis of  elastic $\alpha$ scattering on $^{39}$K reproduced the data over the entire angular range, but it was necessary for the value of the real volume depth to be  greater than 200 MeV. At low energies (around 30 MeV), reduced absorption of the incident $\alpha$- particles is necessary to obtain enhanced elastic angular distributions in the large-angle scattering region. This reduced absorption was explained in terms of angular momentum mismatch between the entrance and exit channels \cite{EBERHARD1970343} and \cite{PLANETA197997}. Another attempt to interpret ALAS was by including a term in the potential that accounts for an $\alpha$ exchange process. A \lq \lq knock-on" (KO) exchange amplitude was calculated using the distorted wave Born approximation \cite{PhysRevC.7.1368}. The introduced KO amplitude peaks at the nuclear surface of the target nucleus, and by semi-classical arguments this term peaks for an angular momentum quantum number $l=kR$, where $k$ is the wave number and $R$ the radius of the target nucleus. For 29 MeV $\alpha$-particles scattered on $^{40}$Ca, the elastic angular distributions in the back-angle scattering region were reproduced by adding the KO amplitude with $l=8$ to the optical model potential \cite{PhysRevC.7.1368}. 

The ALAS effect has been known to be mostly important for $N=Z$ nuclei like: $^{16}$O, $^{20}$Ne, $^{24}$Mg, $^{28}$Si, $^{32}$S, $^{36}$Ar and $^{40}$Ca. However, the availability of radioactive beams (e.g $^{56}$Ni and $^{44}$Ti) made it possible to investigate whether the ALAS effect exists for nuclei beyond $^{40}$Ca. Scattering of $\alpha$-particles on $^{41}$Ca still showed enhanced angular distributions \cite{APELL1975477}, but the ALAS effect seemed to disappear for nuclei heavier than calcium \cite{APELL1975477}. Specifically, in reference \cite{PhysRevLett.89.132501}, the elastic scattering of $\alpha$-particles on the unstable $^{44}$Ti nucleus was considered. The results showed no sign of the ALAS effect. 

Folding potentials were also considered to explain the ALAS effect. The  density-dependent folding model DDM3Y explained the angular distributions for $\alpha$ scattering on $^{13}$C, $^{14, 15}$N and $^{16,17,18}$O but failed to describe the angular distributions for $\alpha$ scattering off the $^{20}$Ne nucleus \cite{Abele1987373} and \cite{PhysRevC.43.2307}. At incident energies  in the 12.7 - 31.1 MeV range, the $\alpha+^{16}$O elastic scattering process was investigated based on the $\alpha+^{16}$O clustering structure of the $^{20}$Ne nucleus. An $\alpha$ folding model, where the imaginary part of the potential was taken to be of the standard Woods-Saxon form, reasonably described the elastic angular distributions \cite{Yang2011655}.  Quasi molecular resonances \cite{RINATREINER1972281} and Regge poles \cite{PhysRevC.3.1104}, molecular type or square Woods-Saxon optical potential \cite{PhysRevC.59.2558},  were also considered as possible explanations for the enhanced, large-angle elastic angular distributions. Despite the many different models that have been used to explain the ALAS effect, so far there has been no one model that could consistently explain the ALAS effect for a wide range of energies and target mass numbers. 

In a previous work, a velocity-dependent optical model (VDOM) was proposed to describe the $p$-wave nature of the pion-nucleus scattering \cite{PhysRev.98.761}. This model gives rise to two terms in the optical potential. One is attractive and was proposed to be a result of a change in mass of the incident nucleon as a consequence of its interactions with the nucleons of the target nucleus. The second is proportional to the derivative of the nuclear matter density of the target. Hence it is most important at the nuclear surface. The strong $l$-dependence of the surface term  enabled the prediction of the large-angle elastic angular distributions for meson scattering off the light $^{12}$C nucleus, which mainly consists of a surface region \cite{PhysRev.98.761}. The VDOM was used to describe neutron elastic scattering off the light $^{12}$C \cite{PhysRevC.84.034618} and intermediate $^{40}$Ca \cite{PhysRevC.85.024606} nuclei. The model was extended to proton scattering off light and intermediate nuclei from $^{12}$C to $^{58}$Ni for incident energies in the range $10 - 40$ MeV \cite{ZUREIKAT2013183}. The VDOM was also extended to higher energies and shown to be effective in describing the large-angle elastic angular distributions for neutron scattering off light and intermediate nuclei \cite{PhysRevC.91.064308}. By fitting nucleon-nucleus ($NA$) elastic angular distributions and polarization data, the potential depths showed linear energy dependence and the geometric parameters were constant for each nuclear target. In the least-square analysis a wide range of incident energies and target nuclei in the mass range $12 \leq A \leq 208$ were used \cite{Alameer_2022}. The success of the VDOM in describing the $NA$ angular distributions even in the large-angle scattering region prompted us to test its effectiveness in describing $\alpha$ scattering off various light, medium and heavy nuclei. The $l$-dependence of the added gradient term that helped in predicting the prominent minima in the large-angle scattering region might play a role in  explaining the ALAS effect. Recently, the VDOM was used to fit the elastic angular distributions for $\alpha$ scattering on the intermediate $^{40}$Ca nucleus. The model reasonably described the enhanced distributions corresponding to incident energies in the range $23 \leq E_{\alpha} \leq 53.9$ MeV \cite{GHABAR2023106335}. 
 
In this work, we shall investigate the large-angle elastic angular distributions corresponding to $\alpha$ scattering off a range of nuclei in the low energy range from 18 to 70 MeV. We shall test the effectiveness of the VDOM  in reproducing the ALAS effect by fitting the elastic angular distributions corresponding to to $\alpha$ scattering off a range of nuclei with mass numbers in the range $12 \leq A \leq 208$ for various energies. In the following section we shall briefly outline the main features of the velocity-dependent optical model. For additional details the reader is referred to reference \cite{PhysRevC.84.034618}.

\section{Velocity-dependent optical model \label{sec:VDP}}        

The conventional optical model (COM)\color{black} was modified by adding to it a gradient term that improved the predictions of the prominent scattering minima that are present  corresponding to back-angle scattering of neutrons on the light $^{12}$C nucleus  \cite{PhysRevC.84.034618}. For a velocity-dependent optical model, \color{black} the Schr\"{o}dinger operator may be expressed in the form:
\begin{equation}
-\mathbf{\nabla }\cdot \frac{\hbar ^{2}}{2m^{\ast }(r)}\mathbf{\nabla +}V-E=-%
\frac{\hbar ^{2}}{2\mu}\mathbf{\nabla }^{2}+\hat{V}(r,p)-E,
\end{equation}
where   $\hat{V}(r,p)$  has the form:
\begin{eqnarray}
\hat{V}(r,p) &=&V(r)+\frac{\hbar ^{2}}{2\mu}\mathbf{\nabla }\cdot \rho (r)%
\mathbf{\nabla },  \nonumber \\
&=&V(r)+\frac{\hbar ^{2}}{2\mu}\left[ \rho (r)\nabla ^{2}+\mathbf{\nabla }%
\rho (r)\cdot \mathbf{\nabla }\right] .  \label{eq:vdp}
\end{eqnarray}
The variable mass $m^*$ was defined to have the following spatial variation Ref.\ \cite{PhysRevC.84.034618}:

\begin{equation}
\frac{1}{m^{\ast }(r)}=\frac{1}{\mu}(1-\rho (r)),  \label{mass}
\end{equation}       
where $\rho(r)$ is an isotropic function of the radial variable $r$. The term $\hat{V}(r,p)$ plays the role of a nonlocal potential $V(r,r')$. This can be seen by interpreting the gradient term $ \left( \mathbf{\nabla }
 \rho (r)\cdot \mathbf{\nabla } \right)$  that acts on the
wave function as the first term of a Taylor series that displaces the wave function from position $\vec{r}$ to
a different location $\vec{r'}$ \cite{FESHBACH1958357}. This gradient term is a surface term, which is proportional to the change in the matter distribution of the nucleus and is most important at the nuclear surface. For light nuclei like $^{12}$C and $^{16}$O that have diffuse edges this term proved to be quite important as such nuclei mainly consist of a surface region. The strong $l$-dependence of this surface term resulted in the prediction of the $p$-wave nature of the pion-nucleus scattering and reproduced the angular distributions at large angles $\theta_{cm} > 90^0$ \cite{PhysRev.98.761}. This same term also resulted in good fits for the elastic angular distributions particularly for the prominent minima in the large-angle scattering region for nucleon scattering off a wide range of nuclei from $^{12}$C to $^{208}$Pb \cite{Alameer_2022}. The nonlocal potential $V(r,r')$ simulates nonlocality, but the origin of this nonlocality is still not quite clear. It could simulate the effect of one or more sources of nonlocality like, channel coupling to inelastic excitations or and the Pauli exclusion principle. \color{black}  

The Schrödinger equation corresponding to the VDOM of equation (\ref{eq:vdp}) has the form:
\begin{equation}
\left( 1 - \rho \right) v''(r) - \left[ v'(r) - \frac{v'(r)}{r}  \right] \rho' - (1-\rho) \frac{l(l+1)}{r^2} v(r)=\frac{2\mu}{\hbar^2} [V(r)-E] v(r), \label{eq:VDPSch}
\end{equation}
where, for clarity, we suppressed the dependencies of $\rho(r)$ on $r$ and of the reduced wave function $v(k,r)$ on the  wave number $k$.

The potential $V(r)$ represents the standard form of the conventional optical model (COM), which consists of the following real volume, imaginary volume, imaginary surface, Coulomb and spin-orbit terms respectively:
\begin{eqnarray}
V(r) &=& -V_{0}f(r,x_{0})- i W_v f(r,x_{0}) + 4ia_{w} W_d \frac{df(r,x_{w})}{dr} + V_C(r) \nonumber \\ 
     &+&  2\left( \frac{\hbar }{%
m_{\pi} c}\right) ^{2}(V_{so}+iW_{so})\frac{1}{r}\frac{df(r,x_{so})}{dr}\vec{%
\sigma}\cdot \vec{I} ,  \label{Vr}
\end{eqnarray}
where $x_{0}$ stands for $(r_{0},a_{0})$ and so on for the rest of the
terms and $m_{\pi}$ is the pion's mass. The function $f(r,r_{j},a_{j})$ has a Woods-Saxon form: 
\begin{equation}
f(r,r_{j},a_{j})=\frac{1}{1+\exp [(r-r_{j}A^{1/3})/a_{j}]},
\end{equation}
with $A$ being the mass number of the target nucleus. \color{black} Clearly, since the spin of the $\alpha$ is zero, there is no need for the spin-orbit terms in this work.  Furthermore, the Coulomb potential is assumed to result from a uniformly charged positive sphere and therefore has the following form:
\begin{eqnarray}
V_C(r) &=& \frac{z Z e^2}{4 \pi \epsilon_0 (2R_C)} \left[ 3 - \left(  \frac{r}{R_C}   \right)^2    \right],  r \leq R_C  \nonumber \\
       &=& \frac{z Z e^2}{4 \pi \epsilon_0 r},  r \geq R_C
\end{eqnarray}
where $R_C=1.20 A^{1/3}$ fm is the Coulomb radius, $z$ and $Z$ are the atomic numbers of the projectile and target respectively. Finally, the isotropic function $\rho(r)$ is assumed to have the following surface and volume forms, respectively; 
\begin{equation}
\rho (r)=\rho_s(r) + \rho_v(r) = \rho _s \; a_{\rho }  \frac{d f(r,x_{\rho})}{dr} + \rho_v  f(r,x_{\rho}).  \label{rho}
\end{equation}
The surface part can be interpreted as the gradient of the mass density of the target nucleus \cite{PhysRev.98.761}, while the volume part is proportional to the target's mass density. At low incident energies, the interactions take place at the nuclear surface, therefore the surface part  $\rho_s(r)$ plays an important role. For incident energies greater than 30 MeV, the volume $\rho_v(r)$ becomes important indicating that interactions inside the nuclear volume become important \cite{ZUREIKAT2013183}.  The above forms of $V(r)$ and $\rho(r)$ resulted in very good fits to the nucleon-nucleus elastic angular distributions \cite{PhysRevC.84.034618}-\cite{Alameer_2022}. Previous works considered the most general form of the kinetic energy operator for a spatially variable mass and its relation to the Schr\"{o}dinger equation for a velocity-dependent optical model \cite{jaghoub2006effect} and \cite{jaghoub2006perturbation}. 

\section{Results and Discussion}
The velocity-dependent optical model has been successful in describing the prominent large-angle minima in the elastic angular distributions for nucleon scattering on light, intermediate and heavy nuclei. Such minima are usually associated with nonlocal effects in the scattering process. As explained in section~\ref{sec:VDP}, the velocity-dependent optical model simulates the role of a nonlocal potential. \color{black} Therefore,  in this work, we shall  asses the the effectiveness of the VDOM in reproducing the enhanced elastic angular distributions for $\alpha$-particle scattering on a range of nuclei with mass numbers in the range $12 \leq A \leq 208$ for various energies in the range $18 \leq E \leq 70$. The nuclei considered in this work together with the corresponding energies are shown in Table~\ref{T:Data}. In what follows we shall discuss the results we obtained corresponding to each of the considered nuclei.

\begin{table}[h]
\caption{List of the nuclei and energies considered in this work for $\alpha$-nucleus elastic scattering process}
\label{T:Data}
\begin{tabular}{@{}lll@{}}
\toprule
Nucleus      & \multicolumn{2}{c}{Elastic Data}                                                                  \\ \midrule
             & $E$(MeV)                                     & Ref.                                               \\ \midrule
${}^{12}$C   & 18.5, 21.7, 25                               & \cite{ignatenko1990experimental}  \\
             & 27.1                                         & \cite{burymov1981study}           \\
             & 48.7, 54.1                                   & \cite{abele1987measurement}       \\
             & 50.5                                         & \cite{pavlova1976direct}          \\
             & 65                                           & \cite{yasue1983deformation}       \\
${}^{16}$O   & 20.1                                         & \cite{ferrero1969new}             \\
             & 25.2, 30.3                                   & \cite{ignatenko1996investigation} \\
             & 39.3, 69.5                                   & \cite{michel1983optical}          \\
             & 48.1                                         & \cite{burtebayev2017scattering}   \\
             & 54.1                                         & \cite{abele1987measurement}       \\
${}^{40}$Ca  & 23, 26.13, 29.05, 31, 34.69, 38.57, 46, 53.9 & \cite{gubler1981phenomenological} \\
             & 49.5                                         & \cite{lohner1978investigation}    \\
${}^{58}$Ni  & 18, 24.1                                     & \cite{trombik1974back}            \\
             & 21.08                                        & \cite{fulmer196821}               \\
             & 25                                           & \cite{ballester1987alpha}         \\
             & 29, 34, 38, 58                               & \cite{budzanowski1978elastic}     \\
             & 32.3                                         & \cite{cowley1974diffraction}      \\
             & 37, 43                                       & \cite{kiebele1979measurement}     \\
${}^{208}$Pb & 27.6                                         & \cite{karcz1971measurements}      \\
             & 39                                           & \cite{gonchar1969alpha}           \\
             & 50                                           & \cite{david1976elastic}           \\
             & 58                                           & \cite{tickle1975208pb}                                    \\ \bottomrule
\end{tabular}
\end{table}

\subsection{$^{12}$C nucleus}
\begin{figure}[htb]
	\centering  
	\includegraphics[scale=0.5]{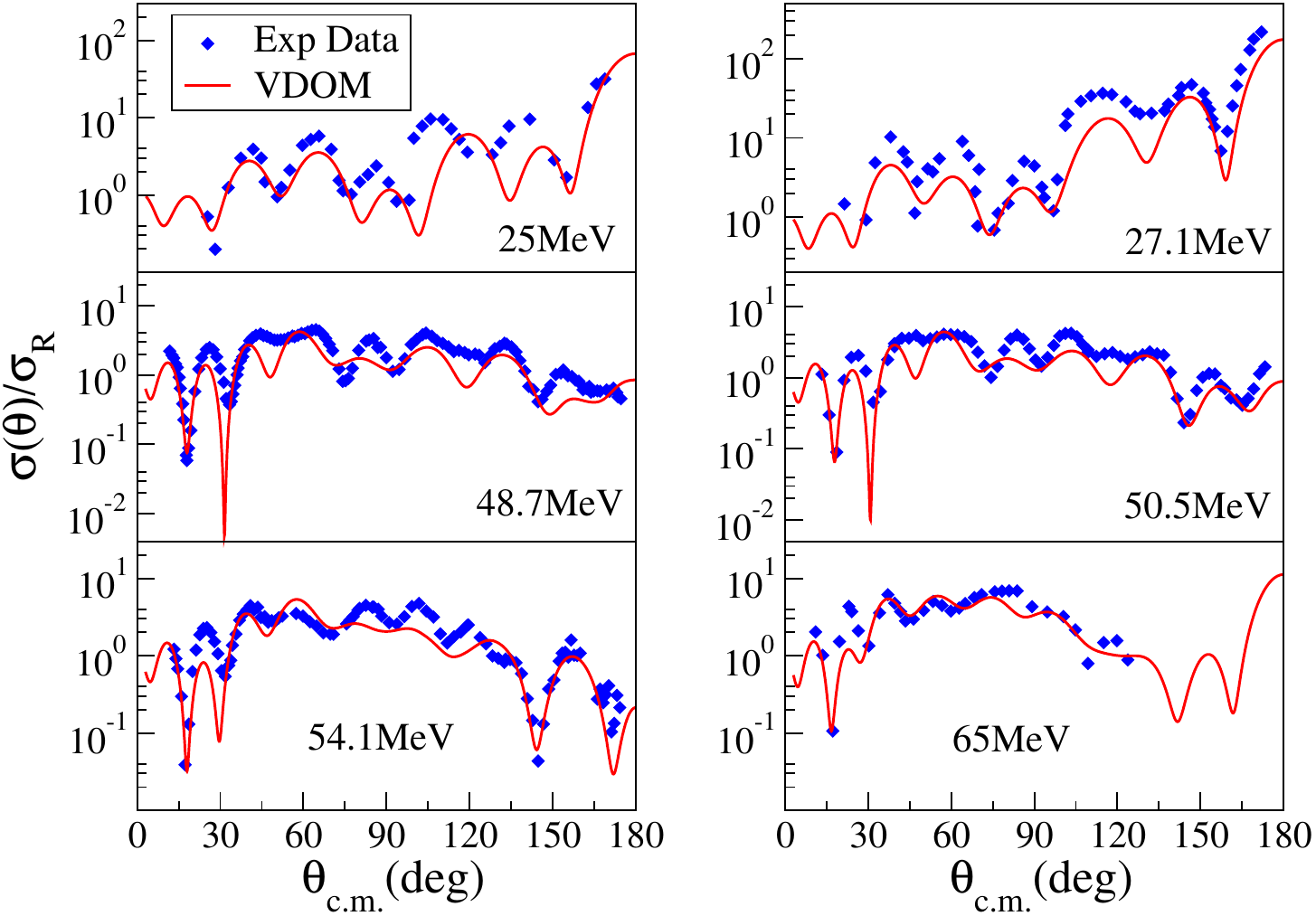}
	\caption{Our best angular distribution fits for $\alpha$ scattering on $^{12}$C obtained using the velocity-dependent optical model (VDOM).  \label{F:C12} }
\end{figure}

We start with the light $1p$-shell $^{12}$C nucleus that is known for its diffuse edge and, therefore, the $\rho(r)$ surface term is expected to play an important role in the scattering process similar to the case of $N-^{12}$C scattering. Inspection of Fig.~\ref{F:C12} shows that the VDOM reproduced the enhanced angular distributions  corresponding to scattering angles greater than $90^{0}$, which are particularly evident for incident energies below  28  MeV energy. It can also be seen that the quality of the fits across the entire angular range improves with increasing incident energy.  The corresponding best-fit parameters are shown in Table~\ref{T:C12-VDP}. In obtaining our parameters we have paid special attention to determine physically meaningful parameters that vary smoothly with energy. This requirement resulted in some increase in the $\chi^2$ values. As shown in Table~\ref{T:C12-VDP relations}, the best-fit parameters  varied linearly with incident energy. To indicate the quality of the linear fit, we have included the correlation coefficient R. \color{black}  The depth of the real volume term ($V$) decreases with energy similar to the case of nucleon-nucleus scattering, but there is a clear change of behavior for energies around 28 MeV. This is likely to be a consequence of the formation of a resonance in the $^{12}$C($\alpha,\alpha)^{12}$C system \cite{CARTER1968202}. In addition, no imaginary volume term was needed for incident energies up to  27 MeV, which indicates that volume absorption does not play a significant role at low energies and absorption mainly takes place at the nuclear surface. At higher energies, however, the $\alpha$-particles have enough energy to overcome the coulomb barrier and volume absorption becomes important as an imaginary volume term is needed. The imaginary surface depth $W_s$ is clearly smaller than the real volume depth and shows a decrease with incident energy. However, similar to the real volume depth ($V$) its linear energy dependence on incident energy below 28 MeV is different to that at higher energies.

\begin{table}[htb]
\caption{Our best-fit velocity-dependent optical model parameters for $\alpha$-particle scattering on the light $^{12}$C nucleus. \color{black}}
\label{T:C12-VDP}
\begin{tabular}{@{}ccccccccc@{}}
\toprule
Parameters & \multicolumn{8}{c}{$E_{\rm lab}$ (MeV)}                                    \\ \midrule
           & 18.5   & 21.7   & 25      & 27.1   & 48.7   & 50.5   & 54.1   & 65     \\ \midrule
$V_v$      & 160.1  & 150.6  & 140.525 & 142.6  & 187.9  & 181.7  & 181.0  & 175.0  \\
$r_v$      & 1.000  & 1.000  & 1.000   & 1.000  & 1.075  & 1.109  & 1.132  & 1.069  \\
$a_v$      & 0.380  & 0.389  & 0.313   & 0.342  & 0.534  & 0.504  & 0.517  & 0.605  \\
$W_v$      & 0    & 0    & 0     & 0    & 4.6    & 5.1    & 4.6    & 4.1    \\
$r_{vi}$   & -      & -      & -       & -      & 1.900  & 1.900  & 1.900  & 1.900  \\
$a_{vi}$   & -      & -      & -       & -      & 1.200  & 1.106  & 1.200  & 1.000  \\
$W_s$      & 20.4   & 20.3   & 19.0    & 11.9   & 22.4   & 20     & 19.4   & 18.0   \\
$r_s$      & 1.000  & 1.058  & 1.131   & 1.152  & 1.096  & 1.103  & 1.138  & 1.006  \\
$a_s$      & 0.200  & 0.200  & 0.200   & 0.200  & 0.200  & 0.200  & 0.200  & 0.200  \\
$\rho_s$   & -2.304 & -2.301 & -2.965  & -2.664 & -1.868 & -1.875 & -1.677 & -1.091 \\
$\rho_v$   & -0.523 & -0.590 & -0.581  & -0.499 & -0.8   & -0.796 & -0.844 & -0.649 \\
$r_{\rho}$ & 1.426  & 1.675  & 1.527   & 1.610  & 1.850  & 1.843  & 1.800  & 1.830  \\
$a_{\rho}$ & 0.500  & 0.500  & 0.500   & 0.500  & 0.500  & 0.500  & 0.5    & 0.500  \\
$\chi^2$   & 49.3   & 36.7   & 25.1    & 26.8   & 18.6   & 15.6   & 19.6   & 10.5   \\ \bottomrule
\end{tabular}
\end{table}

\begin{table}[htb]
\caption{Our potential parameters as linear functions of incident energy for the velocity-dependent optical model adopted in this work corresponding to $\alpha$-particle scattering on the light $^{12}$C nucleus. The corresponding correlation coefficients $R$ are also shown. \color{black}}
\label{T:C12-VDP relations}
\begin{tabular}{@{}lllcc@{}}
\toprule
${}^{12}C$ & $E \leq 27.1$    & $E \geq 48.7$    & \multicolumn{1}{l}{$R$($E \leq 27.1$)} & \multicolumn{1}{l}{$R$ ($E \geq 48.7$)} \\ \midrule
V          & $199.8-2.224E$   & $217.5-0.662E$   & -0.944                                 & -0.918                                  \\
$r_v$      & $0.945+0.00267E$ & $0.945+0.00267E$ & 0.858                                  & 0.858                                   \\
$a_v$      & $0.232+0.00556E$ & $0.232+0.00556E$ & 0.935                                  & 0.935                                   \\
$W_v$      & 0                & $7.11-0.0462E$   & -                                      & -0.800                                  \\
$r_{vi}$   & -                & $1.900$          & -                                      & -                                       \\
$a_{vi}$   & -                & $1.707-0.01064E$ & -                                      & -0.815                                  \\
$W_s$      & $37.79-0.8623E$  & $31.79-0.2168E$  & -0.804                                 & -0.864                                  \\
$r_{s}$    & $0.664+0.01829E$ & $1.422-0.00617E$ & 0.994                                  & -0.801                                  \\
$a_s$      & $0.200$          & $0.200$          & -                                      & -                                       \\
$r_{\rho}$ & $1.384+0.00801E$ & $1.384+0.00801E$ & 0.875                                  & 0.875                                   \\
$a_{\rho}$ & $0.500$          & $0.500$          & -                                      & -                                       \\ \bottomrule
\end{tabular}
\end{table}

\subsection{$^{16}$O nucleus}
\begin{figure}[htb]
	\centering  
	\includegraphics[scale=0.5]{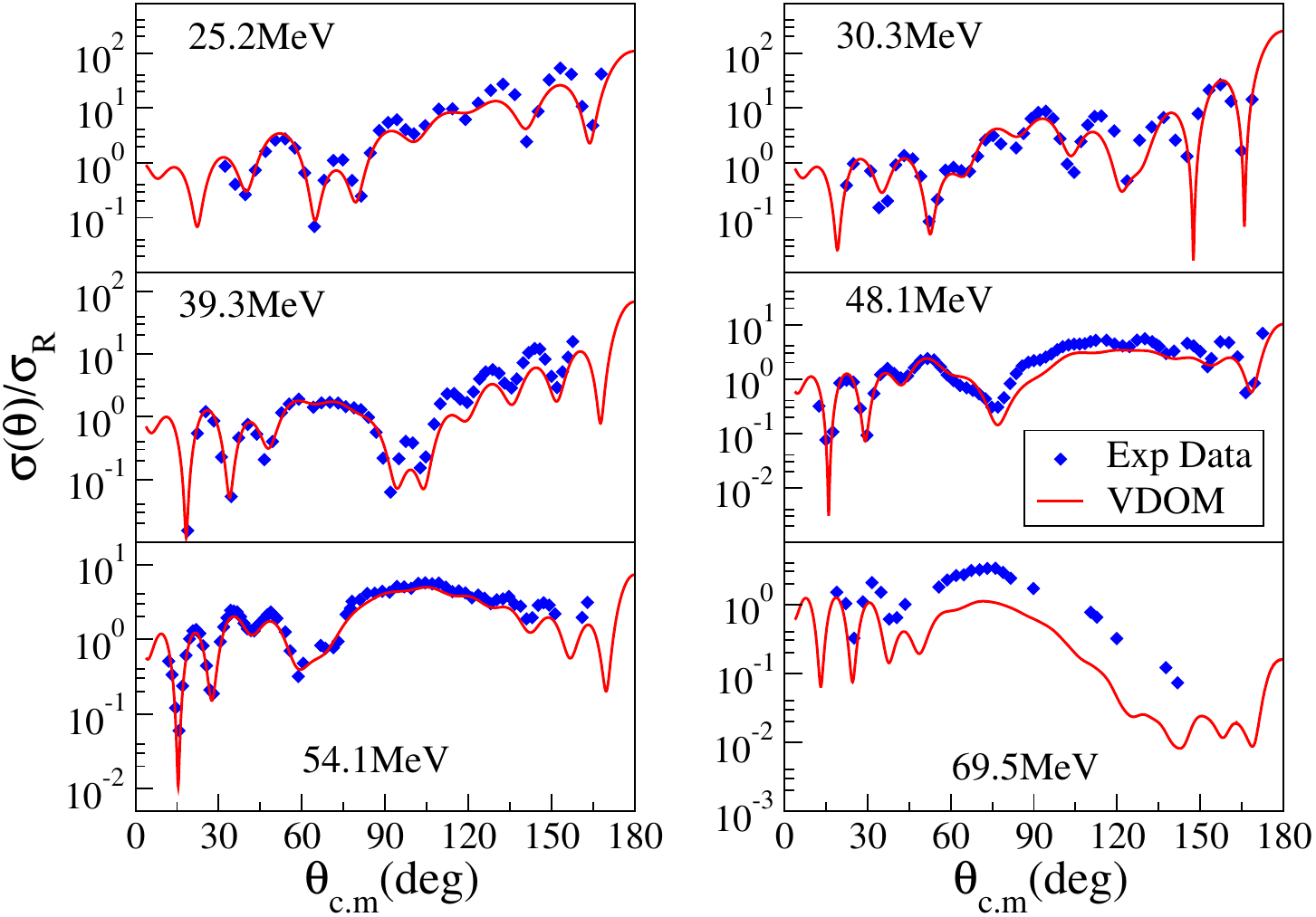}
	\caption{Our best angular distribution fits for $\alpha$ scattering on $^{16}$O obtained using the velocity-dependent optical model (VDOM).  \label{F:O16} }
\end{figure}

Similar to carbon-12, $^{16}$O is a light $1p$-shell nucleus which mainly consists of a surface region and, therefore, the surface gradient term incorporated in the VDOM potential is expected to play an important role in the scattering process. Table~\ref{T:O16-VDP} shows our best-fit parameters obtained using the VDOM, which vary linearly with incident energy as can be seem in Table~\ref{T:O16-VDP Relations}.  Similar to the $^{12}$C case, the values of the potential parameters show changes in their linear behavior for energies above 30 MeV incident energies. For energies below 30 MeV, no imaginary volume term was needed to obtain  our best-fit angular distributions. This is not surprising as the $^{16}$O nucleus mainly consists of a surface region, and volume absorption can occur at higher energies that enable the $\alpha$ particles to overcome the coulomb repulsion barrier and penetrate deep into the nucleus. The corresponding angular distribution fits are compared to the elastic experimental data in Fig.~\ref{F:O16}. Clearly, the VDOM has reproduced the enhanced angular distributions, which are prominent for incident energies below 50 MeV. In addition, the VDM has reproduced the angular distributions across the entire angular range to a very good extent.   
\begin{table}[htb]
\caption{Our best-fit velocity-dependent model parameters for $\alpha$-particle scattering on the light $^{16}$O nucleus.}
\label{T:O16-VDP}
\begin{tabular}{@{}cccccccc@{}}
\toprule
Parameters & \multicolumn{7}{c}{$E_{\rm lab}$ (MeV)}                          \\ \midrule
           & 20.1   & 25.2   & 30.3   & 39.3   & 48.1   & 54.1   & 69.5   \\ \midrule
$V_v$      & 240.1  & 244.6  & 226.9  & 218.4  & 219.6  & 214.8  & 207.1  \\
$r_v$      & 1.028  & 1.000  & 1.020  & 1.012  & 1.043  & 1.074  & 1.064  \\
$a_v$      & 0.723  & 0.686  & 0.777  & 0.638  & 0.634  & 0.644  & 0.538  \\
$W_v$      & 0      & 0      & 18.7   & 23.7   & 23.5   & 27.2   & 25.9   \\
$r_{vi}$   & -      & -      & 1.020  & 1.081  & 1.073  & 1.030  & 1.318  \\
$a_{vi}$   & -      & -      & 0.900  & 0.800  & 1.043  & 1.028  & 1.100  \\
$W_s$      & 8.4    & 24.6   & 0      & 6.6    & 5.6    & 5.6    & 5.3    \\
$r_s$      & 1.508  & 1.381  & -      & 1.260  & 1.360  & 1.271  & 1.243  \\
$a_s$      & 0.200  & 0.200  & -      & 0.200  & 0.200  & 0.200  & 0.200  \\
$\rho_s$   & -2.188 & -1.671 & -1.041 & -1.854 & -1.355 & -1.288 & -1.686 \\
$\rho_v$   & -0.769 & -0.485 & -0.343 & -0.418 & -0.385 & -0.458 & -0.509 \\
$r_{\rho}$ & 1.527  & 1.668  & 1.773  & 1.571  & 1.650  & 1.586  & 1.508  \\
$a_{\rho}$ & 0.516  & 0.500  & 0.500  & 0.523  & 0.567  & 0.532  & 0.562  \\
$\chi^2$   & 53.0   & 15.2   & 30.2   & 21.8   & 14.3   & 4.7    & 45.3   \\ \bottomrule
\end{tabular}
\end{table}

\begin{table}[htb]
\caption{Our potential parameters as linear functions of incident energy for the velocity-dependent optical model adopted in this work corresponding to $\alpha$-particle scattering on the light $^{16}$O nucleus. The corresponding correlation coefficients $R$ are also shown. \color{black}
}
\label{T:O16-VDP Relations}
\begin{tabular}{@{}lllcc@{}}
\toprule
${}^{16}$O & $E < 30.3$       & $E \geq 30.3$    & $R$($E < 30.3$) & $R$ ($E \geq 30.3$) \\ \midrule
$V_v$      & $253.9-0.717E$   & $253.9-0.717E$   & -0.921          & -0.921              \\
$r_v$      & $0.983+0.00125E$ & $0.983+0.00125E$ & 0.802           & 0.802               \\
$a_v$      & $0.815+0.00371E$ & $0.815+0.00371E$ & 0.854           & 0.854               \\
$W_v$      & 0                & $17.19-0.1288E$  & -               & 0.813               \\
$r_{vi}$   & -                & $0.787+0.00657E$ & -               & 0.800               \\
$a_{vi}$   & -                & $0.655-0.00661E$ & -               & 0.809               \\
$W_s$      & $-55.34-3.1712E$ & $4.65-0.0024E$   & 1               & -0.839              \\
$r_{s}$    & $1.525+0.00441E$ & $1.525+0.00441E$ & -0.808          & -0.808              \\
$a_s$      & $0.200$          & $0.200$          & -               & -                   \\
$r_{\rho}$ & $0.971+0.02765E$ & $1.883+0.00551E$ & 1               & -0.817              \\
$a_{\rho}$ & $0.477+0.00125E$ & $0.477+0.00125E$ & 0.806           & 0.806               \\ \bottomrule
\end{tabular}
\end{table}

\subsection{$^{40}$Ca Nucleus}

Many previous works have analyzed the angular distributions corresponding to $\alpha-^{40}$Ca elastic scattering. They all reported the presence of the ALAS effect. One work adopted the following energy-dependent form of the conventional optical model \cite{PhysRevC.18.1237}:

\begin{equation}
U(r,E)=V_C(r) - V_0 f^2(r,d_1,b_1) - iW_v f^2(r,d_2,b_2), 
\end{equation}  
where
\begin{equation}
f(r,d_i,b_i)=\left[ 1 +  \exp\left(\frac{r - d_i A^{1/3}}{b_i}\right) \right]^{-1}
\end{equation}
with $d_1=1.37$ fm, $b_1=1.29$ fm, $d_2=1.75$ fm, $b_2=1.00$ fm, and the energy dependence of the potential is contained in the potential depths, which were assumed to have the following forms:

\begin{eqnarray}
V_0 &=& 198.6 (1 - 0.00168 E)  \label{eq:depths 32} \nonumber \\
W_v &=& 2.98+0.288 E 
\end{eqnarray}
Another work  adopted the following form of the conventional optical model \cite{PhysRevC.16.142}:  

\begin{equation}
U(r)=V_C(r) - V_0 f(r,R_0,s,a) - i W_0 f(r,R_W,s,a), 
\end{equation} 
where
\begin{equation}
f(r,R,s,a) = \left[ 1 +  \exp\left(\frac{r - R}{s\, a}\right) \right]^{-s},
\end{equation}
where
$V_0=-287.9$ fm, $W_0 = -29.9$ MeV, $R_0=4.87$ fm, $a=0.56$, $s=2.65$ and the energy-dependent radius of the imaginary term, namely; 
\begin{equation}
R_W=3.853 + 0.035 E,
\end{equation}
A third more recent work  adopted the VDOM, given in eq.\ (\ref{eq:vdp}), to study the elastic angular distributions for $\alpha$ scattering off $^{40}$Ca .  The least-square fit analysis resulted in the following logarithmic energy-dependence of the potential depths:

\begin{eqnarray}
 		V_0 &=& -5.82 \ln\left(E\right)+193.4 \hspace{0.5cm} {\rm MeV}  \nonumber  \\
 	       W_v &=& 15.56 \ln(E)-35.37 \hspace{0.5cm} {\rm MeV},\nonumber \\
 		W_s &=& 5.80 \ln{\left(E\right)}-16.35 \hspace{0.5cm} {\rm MeV}, \nonumber \\
              \rho_0 &=& 0.421 \ln{\left(E\right)}+2.1288 \label{depths}, 
\end{eqnarray}
with the following constant geometrical parameters: $r_0=1.78$, $a_0=0.770$, $r_v=1.040$, $a_v=0.456$, $r_s=1.700$, $a_s=0.467$, $r_\rho = 1.700$ and $a_\rho = 0.580$ all in units of fermi.

\begin{figure}[htb]
	\centering  
	\includegraphics[scale=0.5]{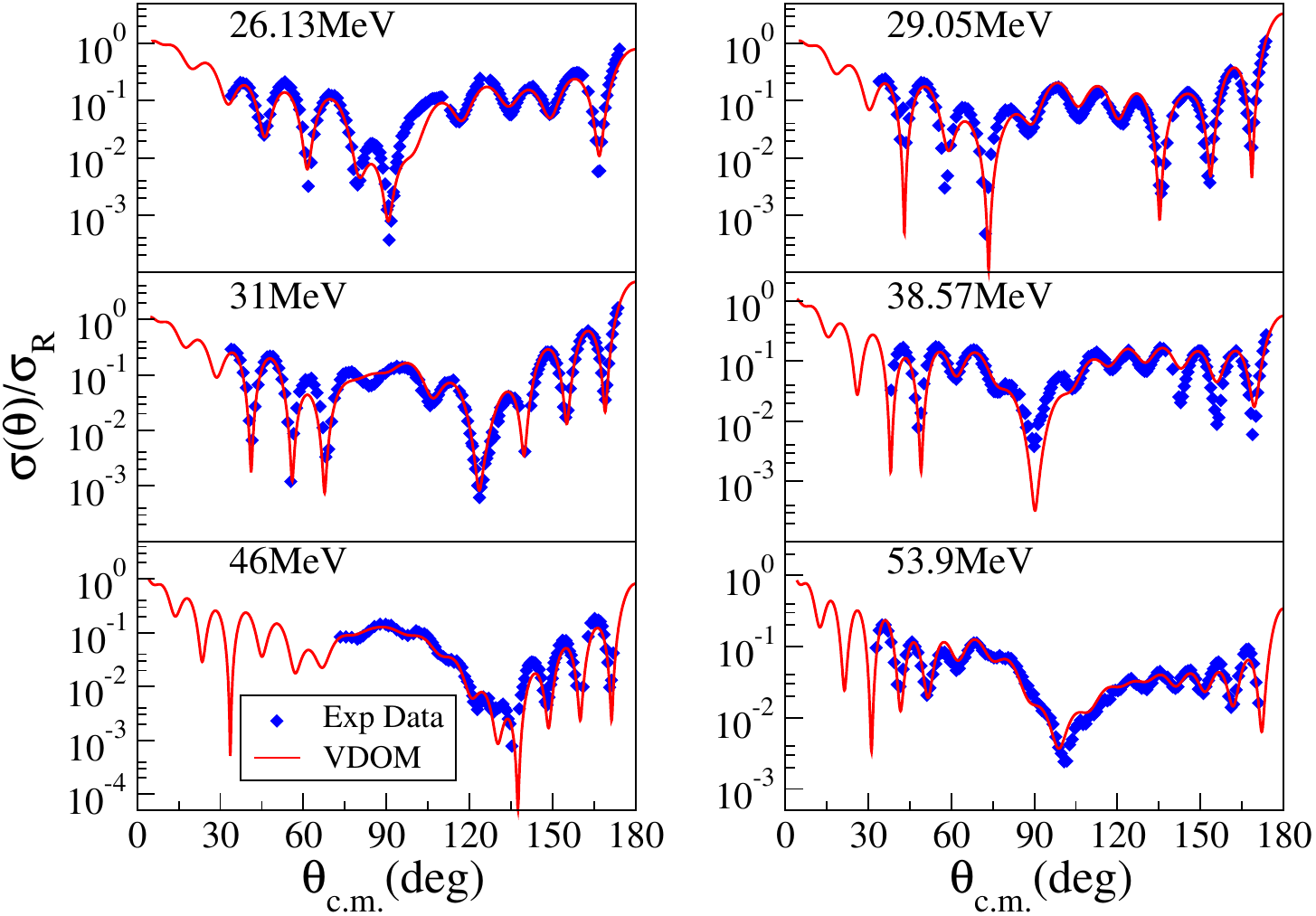}
	\caption{Our best angular distribution fits for $\alpha$ scattering on $^{40}$Ca obtained using the velocity-dependent (VDOM) and conventional optical (COM) models.  \label{F:Ca40} }
\end{figure}

\begin{figure}[htb]
	\centering  
	\includegraphics[scale=0.5]{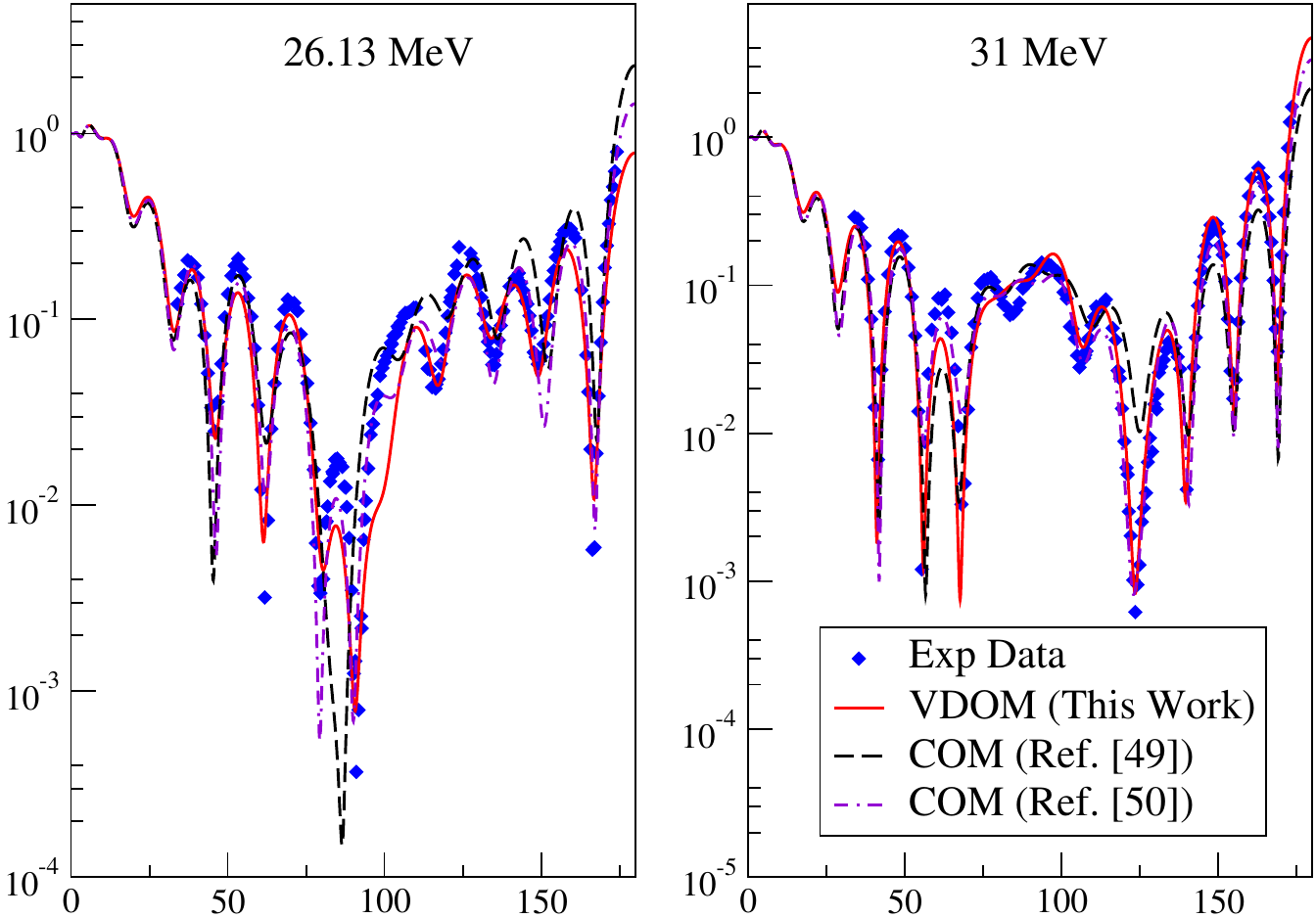}
	\caption{Our best angular distribution fits for $\alpha$ scattering on $^{40}$Ca obtained using the velocity-dependent optical model (VDOM) compared to those obtained using the convention optical model (COM) of the works in Refs.~\cite{PhysRevC.18.1237} and \cite{PhysRevC.16.142}.  \label{F:Ca40-Low-Comp} }
\end{figure}

\begin{figure}[htb]
	\centering  
	\includegraphics[scale=0.5]{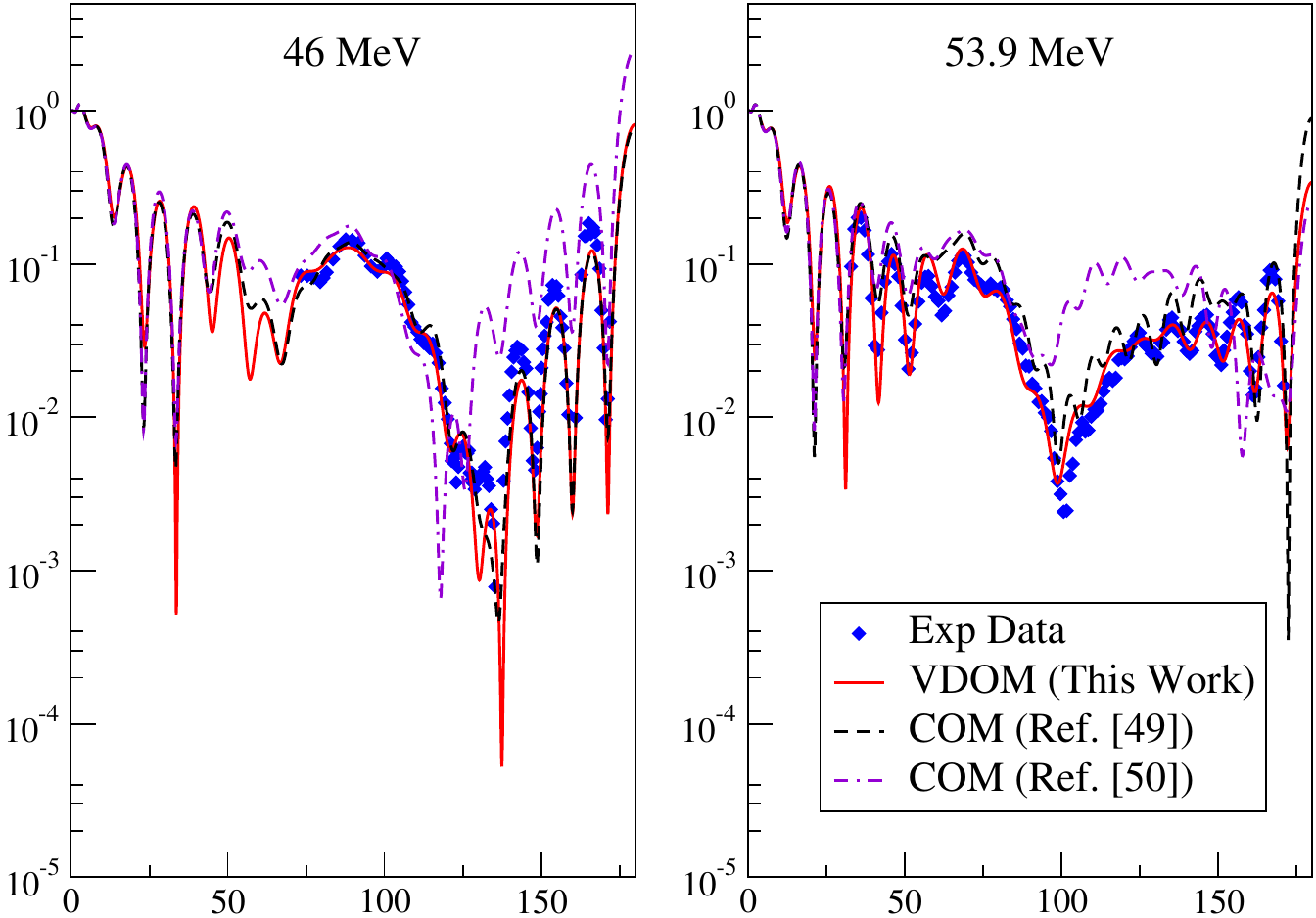}
	\caption{Our best angular distribution fits for $\alpha$ scattering on $^{40}$Ca obtained using the velocity-dependent optical model (VDOM) compared to those obtained using the convention optical model (COM) of the works in Refs.~\cite{PhysRevC.18.1237} and \cite{PhysRevC.16.142}.  \label{F:Ca40-Comp} }
\end{figure}

Inspection of Fig.~\ref{F:Ca40} shows that the VDOM of this work has described the ALAS effect to a very good extent. Our corresponding best-fit parameters are shown in Table~\ref{T:C40-VDP}. The low $\chi^2$ values reflect the good fits to the angular distributions no only in the large angle region but across the full angular range. Unlike the cases of $^{12}$C and $^{16}$O, the imaginary volume term is needed to reproduce the angular distributions even at low incident energies. Table~\ref{T:Ca40-VDP Relations} shows that the potential parameters vary linearly with incident energy.      
To better judge the effectiveness of the velocity-dependent model of this work,  in Table~\ref{T:chi2} we compare our $\chi^2$ values to those calculated using (i) the VDM of Ref.~\cite{GHABAR2023106335}, and (ii) the COM adopted in the works of Refs.~\cite{PhysRevC.18.1237} and \cite{PhysRevC.16.142}. The $\chi^2$ values corresponding to the COM of Ref.~\cite{PhysRevC.18.1237} generally become smaller as the incident energy increases. Contrary to this, the $\chi^2$ values corresponding to the COM of Ref.~\cite{PhysRevC.16.142} become smaller with decreasing incident energy. However, the $\chi^2$ values obtained using the VDOM of this work are small across the entire 23 - 54 considered energy range. The $\chi^2$ values of reference \cite{GHABAR2023106335} are large at the small and large ends of the energy range considered in this work. In fact, the VDOM of this work performs consistently well across the energy range considered in this work. At low incident energies, Fig.~\ref{F:Ca40-Low-Comp} shows that our angular distribution fits and those of Ref.~\cite{PhysRevC.16.142} are better than the ones obtained using the COM of reference \cite{PhysRevC.18.1237}. At higher energies, however, Fig.~\ref{F:Ca40-Comp} shows that our angular distribution fits and those of reference \cite{PhysRevC.18.1237} describe the experimental data better than the model of reference \cite{PhysRevC.16.142}.
\color{black}

\begin{table}[htb]
\caption{Our best-fit velocity-dependent optical model parameters for $\alpha$-particle scattering on the intermediate $^{40}$Ca nucleus. \color{black}}
\label{T:C40-VDP}
\begin{tabular}{@{}cccccccccc@{}}
\toprule
Parameters & \multicolumn{9}{c}{$E_{\rm lab}$ (MeV)}                                        \\ \midrule
           & 23     & 26.13  & 29.05  & 31     & 34.69  & 38.57  & 46     & 49.5   & 53.9   \\ \midrule
$V_v$      & 265.5  & 246.9  & 245.5  & 240.0  & 233.3  & 233.5  & 229.0  & 225.7  & 225.1  \\
$r_v$      & 1.080  & 1.008  & 1.027  & 1.025  & 1.062  & 1.085  & 1.138  & 1.115  & 1.124  \\
$a_v$      & 0.700  & 0.669  & 0.716  & 0.679  & 0.663  & 0.670  & 0.667  & 0.650  & 0.630  \\
$W_v$      & 17.5   & 21.7   & 21.4   & 21.7   & 20.3   & 21.4   & 22.3   & 22.8   & 25.0   \\
$r_{vi}$   & 1.000  & 1.060  & 1.115  & 1.000  & 1.084  & 1.142  & 1.187  & 1.250  & 1.208  \\
$a_{vi}$   & 0.200  & 0.264  & 0.200  & 0.212  & 0.310  & 0.250  & 0.265  & 0.440  & 0.385  \\
$W_s$      & 5.85   & 2.8    & 2.6    & 4.4    & 4.4    & 4.7    & 7.8    & 7.7    & 8.0    \\
$r_s$      & 1.700  & 1.700  & 1.700  & 1.700  & 1.700  & 1.516  & 1.527  & 1.550  & 1.566  \\
$a_s$      & 0.446  & 0.420  & 0.413  & 0.421  & 0.496  & 0.552  & 0.542  & 0.529  & 0.522  \\
$\rho_s$   & -1.576 & -1.852 & -1.473 & -1.691 & -1.549 & -1.554 & -1.156 & -1.751 & -1.520 \\
$\rho_v$   & -0.550 & -0.257 & -0.301 & -0.255 & -0.307 & -0.396 & -0.462 & -0.493 & -0.487 \\
$r_{\rho}$ & 1.690  & 1.658  & 1.630  & 1.679  & 1.649  & 1.537  & 1.576  & 1.503  & 1.492  \\
$a_{\rho}$ & 0.500  & 0.538  & 0.551  & 0.516  & 0.543  & 0.535  & 0.536  & 0.583  & 0.608  \\
$\chi^2$   & 19.4   & 2.5    & 5.3    & 1.5    & 6.8    & 5.3    & 1.8    & 21.2   & 0.5    \\ \bottomrule
\end{tabular}
\end{table}

\begin{table}[htb]
\caption{Our potential parameters as linear functions of incident energy for the velocity-dependent optical model adopted in this work corresponding to $\alpha$-particle scattering on the intermediate $^{40}$Ca nucleus. The corresponding correlation coefficients $R$ are also shown. \color{black}
}
\label{T:Ca40-VDP Relations}
\begin{tabular}{@{}lll@{}}
\toprule
${}^{40}$Ca &                  & $R$    \\ \midrule
V           & $277.5-1.063E$   & -0.893 \\
$r_v$       & $1.160-0.00316E$ & -0.801 \\
$a_v$       & $0.740-0.00187E$ & -0.801 \\
$W_v$       & $16.07+0.1486E$  & 0.800  \\
$r_{vi}$    & $0.846+0.00733E$ & 0.896  \\
$a_{vi}$    & $0.052+0.00621E$ & 0.804  \\
$W_s$       & $0.610+0.1072E$   & 0.802  \\
$r_{s}$     & $1.866+0.00643E$ & -0.816 \\
$a_s$       & $0.327+0.00421E$ & 0.800  \\
$r_{\rho}$   & $1.839-0.00643E$ & -0.915 \\
$a_{\rho}$   & $0.456+0.00242$  & 0.803  \\ \bottomrule
\end{tabular}
\end{table}

\begin{table}[htb]
\caption{Our $\chi^2$ values calculated using the velocity-dependent model for the $\alpha-^{40}$Ca elastic scattering compared the corresponding values calculated using the velocity-dependent model of Ref.~\cite{GHABAR2023106335} and the conventional optical models of Refs.~\cite{PhysRevC.18.1237} and \cite{PhysRevC.16.142}.\color{black}}
\label{T:chi2}
\begin{tabular}{@{}llllll@{}}
\toprule
$E$   &  & $\chi^2$       &                                                   &                                                   &                                                  \\ \midrule
MeV   &  & VDOM(This Work) & VDOM(Ref.\cite{GHABAR2023106335}) & COM(Ref.~\cite{PhysRevC.18.1237}) & COM(Ref.~\cite{PhysRevC.16.142}) \\ \midrule
23    &  & 19.4           & 2541.5                                            & 45.81                                             & 6.95                                             \\
26.13 &  & 2.5            & 251.4                                             & 57.96                                             & 3.98                                             \\
29.05 &  & 5.3            & 40.5                                              & 34.35                                             & 6.84                                             \\
31    &  & 1.5            & 84.4                                              & 31.31                                             & 4.15                                             \\
34.69 &  & 6.8            & 14.9                                              & 5.75                                              & 13.00                                            \\
38.57 &  & 5.3            & 9.2                                               & 3.70                                              & 8.30                                             \\
46    &  & 1.8            & 16.2                                              & 1.49                                              & 245.06                                           \\
49.5  &  & 21.2           & 579.2                                             & 37.71                                             & 3443.87                                          \\
53.9  &  & 0.5            & 116.5                                             & 7.90                                              & 141.22                                           \\ \bottomrule
\end{tabular}
\end{table}

\subsection{$^{58}$Ni Nucleus}

\begin{figure}[htb]
	\centering  
	\includegraphics[scale=0.5]{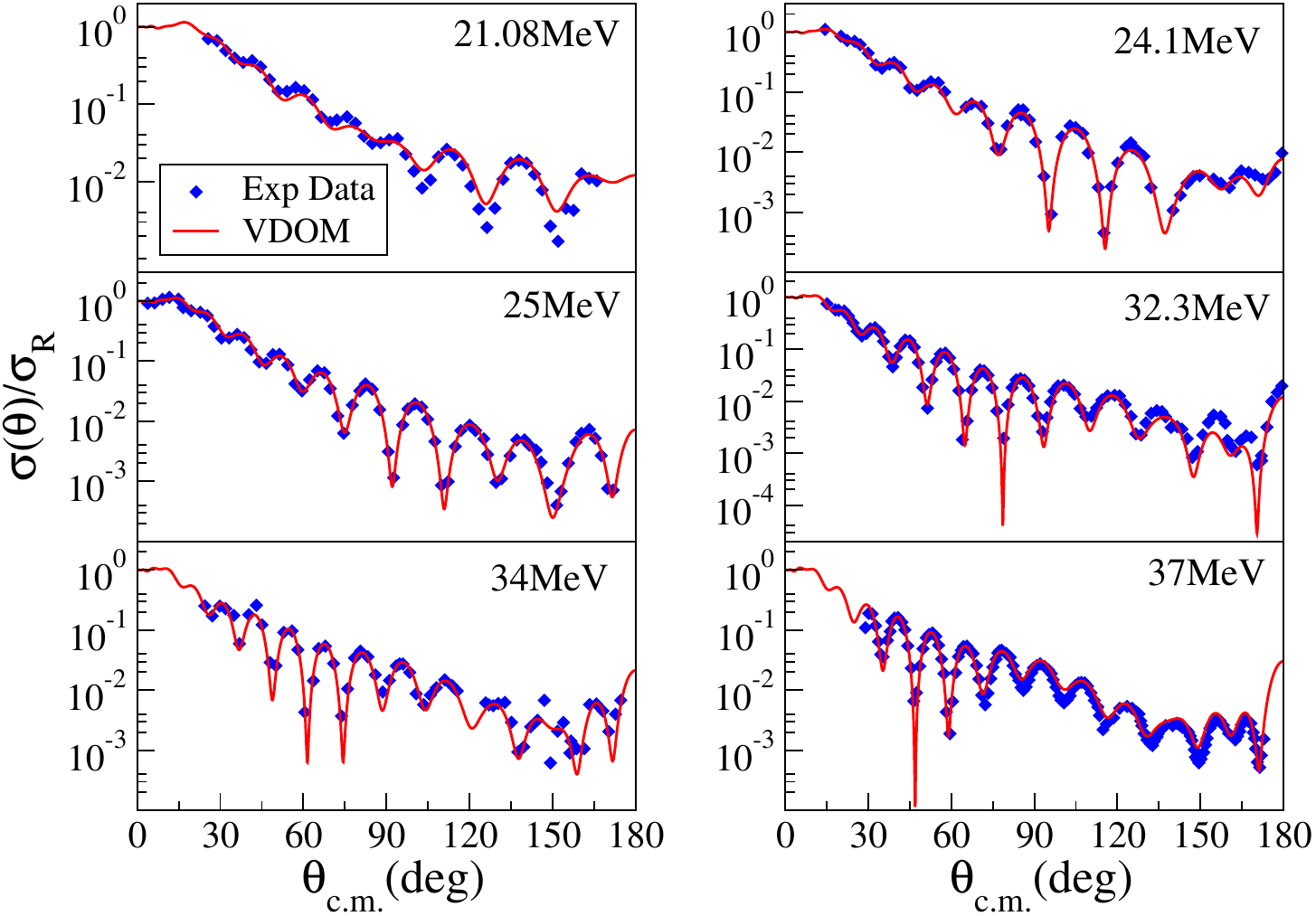}
	\caption{Our best angular distribution fits for $\alpha$ scattering on $^{58}$Ni obtained using the velocity-dependent optical model (VDOM).  \label{F:Ni58} }
\end{figure}

Inspection of Fig.~\ref{F:Ni58} shows that the observed ALAS effect in light nuclei up to $^{40}$Ca is not  present in the case of $\alpha$ scattering on the intermediate $^{58}$Ni nucleus. This is inline with the conclusion of Ref.~\cite{APELL1975477}, which states that the ALAS effect seems to disappear for nuclei heavier than $^{41}$Ca. The $\chi^2$ values in Table~\ref{T:Ni-VDP}  show that the VDOM reproduced the $\alpha - ^{58}$Ni elastic angular distributions well as can be seen in Fig.~\ref{F:Ni58}. Inspection of Table~\ref{T:Ni58-VDP Relations} shows that the potential parameters depend linearly on the energy of the incident $\alpha$ particles.

\begin{table}[htb]
\caption{Our best-fit velocity-dependent optical model parameters for $\alpha$-particle scattering on the intermediate $^{58}$Ni nucleus. \color{black}}
\label{T:Ni-VDP}
\begin{tabular}{@{}cccccccccccc@{}}
\toprule
Parameters & \multicolumn{11}{c}{$E_{\rm lab}$ (MeV)}                                                         \\ \midrule
           & 18     & 21.08  & 24.1   & 25     & 29     & 32.3   & 34     & 37     & 38     & 43     & 58     \\ \midrule
$V_v$      & 136.3  & 126.9  & 118.2  & 119.4  & 117.6  & 118.9  & 118.5  & 117.2  & 110.0  & 103.2  & 92.4   \\
$r_v$      & 1.140  & 1.097  & 1.089  & 1.094  & 1.068  & 1.000  & 1.024  & 1.023  & 1.040  & 1.047  & 1.000  \\
$a_v$      & 0.76   & 0.808  & 0.733  & 0.764  & 0.758  & 0.765  & 0.751  & 0.700  & 0.692  & 0.676  & 0.681  \\
$W_v$      & 13.3   & 13.3   & 13.6   & 15.1   & 13.2   & 14.6   & 14.6   & 15     & 13.9   & 15.5   & 16.1   \\
$r_{vi}$   & 1.460  & 1.400  & 1.394  & 1.409  & 1.420  & 1.306  & 1.370  & 1.313  & 1.360  & 1.368  & 1.280  \\
$a_{vi}$   & 0.200  & 0.200  & 0.200  & 0.229  & 0.200  & 0.248  & 0.284  & 0.435  & 0.386  & 0.300  & 0.410  \\
$W_s$      & 3.8    & 3.4    & 2.8    & 3.0    & 3.7    & 4.4    & 4.2    & 4.1    & 3.0    & 4.4    & 8.4    \\
$r_s$      & 1.700  & 1.700  & 1.685  & 1.677  & 1.653  & 1.642  & 1.654  & 1.618  & 1.561  & 1.542  & 1.592  \\
$a_s$      & 0.381  & 0.387  & 0.357  & 0.385  & 0.410  & 0.465  & 0.460  & 0.459  & 0.537  & 0.635  & 0.535  \\
$\rho_s$   & -2.443 & -2.203 & -2.639 & -2.488 & -2.634 & -2.007 & -2.093 & -1.982 & -1.868 & -2.092 & -1.865 \\
$\rho_v$   & -0.310 & -0.432 & -0.357 & -0.382 & -0.345 & -0.269 & -0.286 & -0.267 & -0.307 & -0.337 & -0.333 \\
$r_{\rho}$ & 1.614  & 1.570  & 1.589  & 1.582  & 1.595  & 1.522  & 1.572  & 1.545  & 1.514  & 1.466  & 1.475  \\
$a_{\rho}$ & 0.500  & 0.500  & 0.500  & 0.516  & 0.500  & 0.590  & 0.539  & 0.581  & 0.524  & 0.577  & 0.643  \\
$\chi^2$   & 23.7   & 8.6    & 3.2    & 2.9    & 35.4   & 8.9    & 3.9    & 13.5   & 33.6   & 29.6   & 32.6   \\ \bottomrule
\end{tabular}
\end{table}

\begin{table}[htb]
\caption{Our potential parameters as linear functions of incident energy for the velocity-dependent optical model adopted in this work corresponding to $\alpha$-particle scattering on the intermediate $^{58}$Ni nucleus. The corresponding correlation coefficients $R$ are also shown. \color{black}}
\label{T:Ni58-VDP Relations}
\begin{tabular}{@{}lll@{}}
\toprule
${}^{58}$Ni & \multicolumn{1}{l}{} & \multicolumn{1}{l}{$R$} \\ \midrule
V           & $147.3-0.953E$       & -0.941                  \\
$r_v$       & $1.160-0.00316E$     & -0.801                  \\
$a_v$       & $0.833-0.00298E$     & -0.800                  \\
$W_v$       & $12.10+0.0697E$      & 0.800                   \\
$r_{vi}$    & $1.496-0.00382E$     & -0.804                  \\
$a_{vi}$    & $0.073+0.00636E$     & 0.800                   \\
$W_s$       & $0.610+0.1072E$       & 0.802                   \\
$r_{s}$     & $0.664+0.01829E$     & -0.801                  \\
$a_s$       & $0.259+0.00600E$     & 0.803                   \\
$r_{\rho}$  & $1.672-0.00376E$     & -0.866                  \\
$a_{\rho}$  & $0.422+0.00370$      & 0.872                   \\ \bottomrule
\end{tabular}
\end{table}

\subsection{The $^{208}$Pb Nucleus} 

\begin{figure}[htb]
	\centering  
	\includegraphics[scale=0.5]{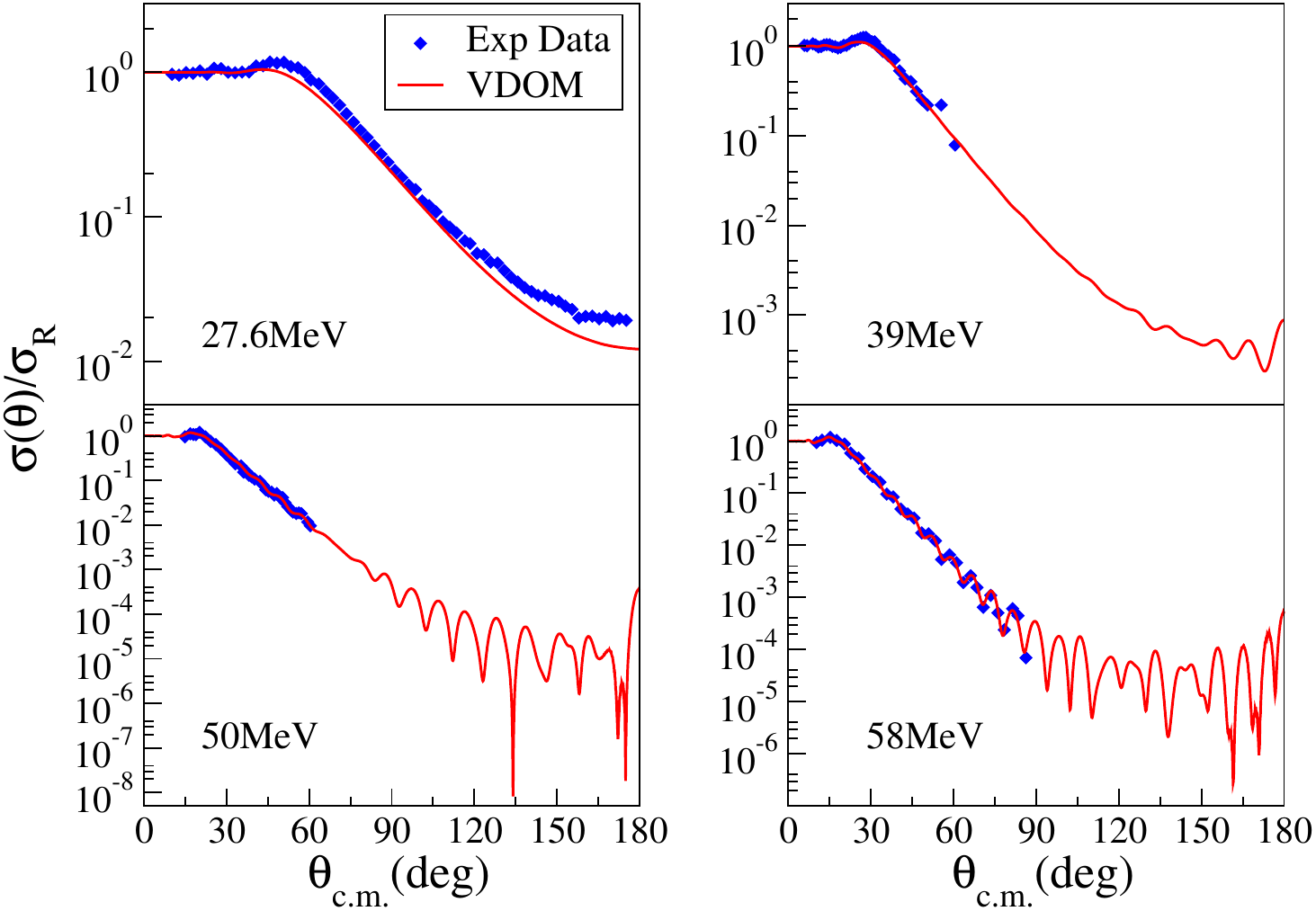}
	\caption{Our best angular distribution fits for $\alpha$ scattering on $^{208}$Pb obtained using the velocity-dependent optical model (VDOM).  \label{F:Pb208} }
\end{figure}

Figure~\ref{F:Pb208} clearly shows that the ALAS effect is not present in the case of $\alpha$ scattering on the heavy $^{208}$Pb nucleus. Still, we included this case to show the effectiveness of the VDOM  in reproducing the experimental angular distributions corresponding to $\alpha$ scattering off heavy nuclei. Table~\ref{T:Pb-VDP} shows our best-fit parameters. The small $\chi^2$ values show that the VDOM reproduced the elastic angular distributions to a very good extent.In addition, Table~\ref{T:Pb208-VDP Relations} shows that the best-fit parameters depend linearly on the incident energy.

\begin{table}[htb]
\caption{Our best-fit velocity-dependent optical model parameters for $\alpha$-particle scattering on the heavy $^{208}$Pb nucleus. \color{black}}
\label{T:Pb-VDP}
\begin{tabular}{@{}ccccc@{}}
\toprule
Parameters & \multicolumn{4}{c}{$E_{\rm lab}$ (MeV)} \\ \midrule
           & 27.6     & 39       & 50      & 58      \\ \midrule
$V_v$      & 180.0    & 173.0    & 166.5   & 161.7   \\
$r_v$      & 1.021    & 1.100    & 1.195   & 1.134   \\
$a_v$      & 0.700    & 0.750    & 0.789   & 0.760   \\
$W_v$      & 12.0     & 14.0     & 16.1    & 19.4    \\
$r_{vi}$   & 1.379    & 1.371    & 1.279   & 1.305   \\
$a_{vi}$   & 0.200    & 0.200    & 0.200   & 0.200   \\
$W_s$      & 7.00     & 6.50     & 6.50    & 5.10     \\
$r_s$      & 1.450    & 1.460    & 1.438   & 1.422   \\
$a_s$      & 0.820    & 0.766    & 0.782   & 0.759   \\
$\rho_s$   & -2.029   & -2.440   & -2.465  & -2.698  \\
$\rho_v$   & -0.298   & -0.527   & -0.504  & -0.347  \\
$r_{\rho}$ & 1.459    & 1.4      & 1.374   & 1.339   \\
$a_{\rho}$ & 0.712    & 0.600    & 0.550   & 0.573   \\
$\chi^2$   & 3.7      & 0.8      & 0.5     & 0.5     \\ \bottomrule
\end{tabular}
\end{table}

\begin{table}[htb]
\caption{Our potential parameters as linear functions of incident energy for the velocity-dependent optical model adopted in this work corresponding to $\alpha$-particle scattering on the heavy $^{208}$Pb nucleus. The corresponding correlation coefficients $R$ are also shown. \color{black}}
\label{T:Pb208-VDP Relations}
\begin{tabular}{@{}lll@{}}
\toprule
${}^{208}$Pb &                  & $R$    \\ \midrule
V            & $196.5-0.600E$   & -1.000 \\
$r_v$        & $0.917-0.00449E$ & 0.819  \\
$a_v$        & $0.651-0.00227E$ & 0.811  \\
$W_v$        & $5.12+0.2352E$   & 0.979  \\
$r_{vi}$     & $1.471+0.00316E$ & -0.850 \\
$a_{vi}$     & $0.200$          & -      \\
$W_s$        & $8.63+0.0539E$   & -0.854 \\
$r_{s}$      & $1.866+0.00643E$ & -0.816 \\
$a_s$        & $0.854+0.00165E$ & -0.800 \\
$r_{rho}$    & $1.558-0.00378E$ & -0.989 \\
$a_{rho}$    & $0.817-0.00476$  & -0.878 \\ \bottomrule
\end{tabular}
\end{table}

\section{Conclusions}
In this work we have tested the effectiveness of the velocity-dependent optical model in describing the ALAS effect corresponding to $\alpha$ scattering on various nuclei from $^{12}$C to $^{40}$Ni. The VDOM has reproduced the enhanced angular distributions (ALAS effect) to a very good extent. Although the ALAS effect is not observed for heavier nuclei, in the least-square fit analysis, we included the intermediate $^{58}$Ni and heavy $^{208}$Pb nuclei to show the effectiveness of the VDOM in describing the elastic angular distributions. For light, intermediate and heavy nuclei, the calculated angular distributions compared well to the experimental data across the entire energy range.  Our best-fit potential parameters are simple linear functions of incident energy. The angular momentum dependence of the gradient term in the VDOM may be responsible for the effectiveness of the model in describing the ALAS effect. 

For $\alpha$ scattering on $^{40}$Ca, we compared our results to those of references \cite{PhysRevC.18.1237} and \cite{PhysRevC.16.142} that adopted the conventional optical model. The first described the angular distributions at higher energies better than at lower ones. In contrast, the latter one described the data better at lower energies. The VDOM of this work, however, described the experimental data to a very good extent for all the energies considered in the range 18 - 70 MeV. We also compared our $\chi^2$ values  to those obtained using the work of reference \cite{GHABAR2023106335} that employed a velocity-dependent optical model. Our values are smaller for all the considered energies.     
\color{black}

\end{document}